\documentclass[reprint,onecolumn,superscriptaddress,amsmath,amssymb,aps,pre,longbibliography,nofootinbib]{revtex4-2}

\usepackage[caption=false,position=top]{subfig}
\usepackage{enumitem}
\usepackage{amsmath}
\usepackage{mathtools}
\usepackage{graphicx}
\usepackage{bm}
\usepackage[dvipsnames]{xcolor}
\usepackage[colorlinks=true,allcolors=red!50!black]{hyperref}
\usepackage{microtype}
\usepackage{svg}

\DeclareMathOperator{\erf}{erf}

\begin{document}

\title{Nadaraya–Watson kernel smoothing as a random energy model}

\author{Jacob A. Zavatone-Veth}
\email{jzavatoneveth@fas.harvard.edu}
\affiliation{Society of Fellows, Harvard University, Cambridge, MA, USA}
\affiliation{Center for Brain Science, Harvard University, Cambridge, MA, USA}

\author{Cengiz Pehlevan}
\email{cpehlevan@seas.harvard.edu}
\affiliation{Center for Brain Science, Harvard University, Cambridge, MA, USA}
\affiliation{John A. Paulson School of Engineering and Applied Sciences, Harvard University, Cambridge, MA, USA}
\affiliation{Kempner Institute for the Study of Natural and Artificial Intelligence, Harvard University, Cambridge, MA, USA}

\begin{abstract}
    Precise asymptotics have revealed many surprises in high-dimensional regression. These advances, however, have not extended to perhaps the simplest estimator: direct Nadaraya-Watson (NW) kernel smoothing. Here, we describe how one can use ideas from the analysis of the random energy model (REM) in statistical physics to compute sharp asymptotics for the NW estimator when the sample size is exponential in the dimension. As a simple starting point for investigation, we focus on the case in which one aims to estimate a single-index target function using a radial basis function kernel on the sphere. Our main result is a pointwise asymptotic for the NW predictor, showing that it re-scales the argument of the true link function. Our work provides a first step towards a detailed understanding of kernel smoothing in high dimensions. 
\end{abstract}

\date{\today}

\maketitle

\section{Introduction}

Assume one has some scalar function $f(x)$ of a $d$-dimensional vector $x$, which one wants to estimate given access to $n$ examples $\mathcal{D}=\{(x_{\mu}, f(x_{\mu})\}_{\mu=1}^{n}$. A classic approach to this problem is \emph{Nadaraya–Watson kernel smoothing} (henceforth the NW estimator), which given a choice of kernel $k(x,x')$ produces the estimate
\begin{align} \label{eqn:nw_estimator}
    \hat{f}_{\mathcal{D}}(x) = \frac{\sum_{\mu=1}^{n} k(x,x_{\mu}) f(x_{\mu})}{\sum_{\mu=1}^{n} k(x,x_{\mu})} . 
\end{align}
The key question is then how close $\hat{f}_{\mathcal{D}}(x)$ is to the true function $f(x)$. One might measure this in terms of the mean-squared error, averaged over realizations of the training set, $\mathbb{E}_{\mathcal{D}} \mathbb{E}_{x}[ (f(x) - \hat{f}_{\mathcal{D}}(x))^2 ]$, or perhaps in terms of a worst-case error $\sup_{x} |f(x) - \hat{f}_{\mathcal{D}}(x)|$. 

Classical results show that the NW estimator suffers from the curse of dimensionality in the sense that $n$ must be exponentially large in the dimension $d$ to achieve error below some fixed tolerance \cite{conn2019oracle,belkin2019interpolation,tsybakov2008nonparam}. This contrasts with the closely-related kernel ridge regression (KRR) estimator, 
\begin{align}
    \hat{f}_{\mathcal{D}}^{\mathrm{KRR}}(x) = \sum_{\mu,\nu=1}^{n} k(x,x_{\mu}) (K^{-1})_{\mu\nu} f(x_{\nu}), \qquad K_{\mu\nu} = k(x_{\mu},x_{\nu}),
\end{align}
which depending on the details of the target function $f$ may require only polynomially many samples to achieve a given accuracy \cite{belkin2019interpolation,canatar2021spectral,xiao2022precise,atanasov2024scaling,spigler2020asymptotic}. This separation is striking given that both of these methods linearly smooth the training labels $f(x_{\mu})$; as highlighted by \citet{belkin2019interpolation}, the key difference is that the NW estimator is a `direct' smoother, while KRR estimates an inverse model for the training data. Recent years have seen substantial advances in understanding the high-dimensional asymptotics of KRR \cite{canatar2021spectral,xiao2022precise,atanasov2024scaling,belkin2019interpolation}, but a similarly detailed understanding of the NW estimator is lacking \cite{belkin2019interpolation,simon2024onenn}. In particular, what are the physical reasons for the poor performance of direct smoothing in high dimensions relative to building an inverse model? 

In this note, we explore an approach to characterizing the NW estimator based on the observation that it can be interpreted in terms of a Random Energy Model (REM) \cite{mezard2009information,lucibello2024exponential,derrida1981rem}.\footnote{See \citet{mezard2009information} for a pedagogical introduction to the REM.} Concretely, we can view \eqref{eqn:nw_estimator} as the average of the observable $f(x_{\mu})$ with respect to a distribution over $\mu$ with Boltzmann weights $k(x,x_{\mu})$ that have quenched randomness due to the randomness of the datapoints $x_{\mu}$. For suitable choices of the data distribution, kernel, and target function, it should thus be possible to apply standard approaches to the REM to study the high-dimensional behavior of the NW estimator. The slightly peculiar feature of the problem is that both the energies and the observable depend on the quenched disorder, \emph{i.e.}, on the random data. From the REM perspective, the relevant regime is when $n = e^{\alpha d}$ as $d \to \infty$ for some fixed $\alpha > 0$, which would match the sample complexity required by the curse of dimensionality. Moreover, one expects that under suitable conditions the REM associated to the estimator will display a condensation transition: there will be a phase in which all datapoints contribute relatively equally to the estimate, and another in which the distribution is sharply condensed, and only a few points contribute. 

\section{Large deviations analysis for spherical data}\label{sec:large_deviations}

We now leverage the REM analogy to analyze the asymptotic behavior of $\hat{f}_{\mathcal{D}}(x)$. For simplicity, we fix a concrete model for the data, kernel, and target function, though many parts of this analysis can be generalized. We assume that the inputs lie on the sphere of radius $\sqrt{d}$, with a fixed test point $x \in \mathbb{S}^{d-1}(\sqrt{d})$ and uniformly-drawn training examples $x_{\mu} \sim \mathcal{U}[\mathbb{S}^{d-1}(\sqrt{d})]$, and take the kernel to be a radial basis function on the sphere, \emph{i.e.}, 
\begin{align}
    k(x,x_{\mu}) = e^{\beta \langle x, x_{\mu} \rangle }
\end{align}
for some inverse bandwidth $\beta>0$. This makes the mapping to the REM quite direct, and is particularly closely related to \citet{lucibello2024exponential}'s recent work on dense associative memories (DAMs) \cite{krotov2016dense}, where it is shown that the memorization capacity for spherical patterns is determined by the free energy of an auxiliary REM. 

With these choices, we focus on the simplest class of functions on the sphere with underlying low-dimensional structure: single index models. Such functions take the form
\begin{align}
    f(x) = g({\langle w, x \rangle}/{d} )
\end{align}
for a vector $w \in \mathbb{S}^{d-1}(\sqrt{d})$ and a scalar link function $g: \mathbb{R} \to \mathbb{R}$. For now, we will assume that the function $g$ is independent of the dimension $d$; we relax that assumption in Section \ref{sec:large_target}. We will often consider positive-homogeneous link functions satisfying $g(\gamma t) = \gamma^{k} g(t)$ for all $\gamma > 0$ and some degree $k \geq 0$; among these is of course the linear rectifier $g(t) = \max\{0,t\}$. 

We begin by considering the predictions on some fixed test point $x \in \mathbb{S}^{d-1}(\sqrt{d})$. For dimension-independent link functions, we will argue that there is an explicitly computable $r_{\ast} = r_{\ast}(\alpha,\beta)$, bounded from below by zero and from above by 1, such that we have the pointwise asymptotic
\begin{align}
    \hat{f}_{\mathcal{D}}(x) \sim g( r_{\ast} \langle w, x\rangle/d ) .
\end{align}
To obtain this result, we start with the fact that, under the stated assumptions, \eqref{eqn:nw_estimator} becomes
\begin{align}\label{eqn:spherical_nw}
    \hat{f}_{\mathcal{D}}(x) = \frac{\sum_{\mu=1}^{n} e^{\beta \langle x, x_{\mu}\rangle} g(\langle w, x_{\mu}\rangle/d)}{\sum_{\mu=1}^{n}  e^{\beta \langle x, x_{\mu}\rangle}} .
\end{align}
By symmetry, the distribution of $\hat{f}_{\mathcal{D}}(x)$ induced by the randomness in the training data can depend on $x$ and $w$ only through their overlap 
\begin{align}
    \rho = {\langle w,x\rangle}/{d}.
\end{align}

Based on standard REM results, we expect the empirical joint distribution of the overlaps 
\begin{align}
    t = \langle x, x_{\mu} \rangle/d 
    \quad \mathrm{and} \quad
    q = \langle w, x_{\mu} \rangle /d, 
\end{align}
viewed as random variables with randomness induced by the empirical distribution of $x_{\mu}$ over $\mu$, to satisfy a large deviation principle. Concretely, what we expect is that for $n = e^{\alpha d}$ and $d \to \infty$, there is a potential $\phi(t,q)$ such that 
\begin{align}
    \hat{f}_{\mathcal{D}}(x) \sim \frac{\int dt\, dq\, e^{d \phi(t,q)} g(q)}{\int dt\, dq\, e^{d \phi(t,q)}} .
\end{align}
This potential will consist of an energetic contribution $\beta t$ from the kernel, plus an entropic contribution resulting from re-writing the sum \eqref{eqn:spherical_nw} in terms of the density of states with a given $(t,q)$. 
We will follow the standard random energy model analysis to obtain an asymptotic of this form, closely following the related work of \citet{lucibello2024exponential}.\footnote{As we will invoke results from \cite{lucibello2024exponential} without proof, we provide a brief dictionary of notation in Appendix \ref{sec:notation}.} Pedagogical accounts of this analysis can also be found in the textbook of \citet{mezard2009information}, or in the original paper of \citet{derrida1981rem}. 

For the purpose of the computation, it is more convenient to instead work in coordinates
\begin{align}
    u = (t+q)/\sqrt{2}
    \quad \mathrm{and} \quad
    v = (t-q)/\sqrt{2},
\end{align}
as by the Cauchy-Schwarz inequality these variables lie within the ellipse $u^2/(1+\rho) + v^2/(1-\rho) \leq 1$. The first step is to determine the density of states with a given $u$ and $v$, leveraging the G\"artner-Ellis theorem. Following \citet{lucibello2024exponential}, it is easy to show that the joint moment generating function of $u$ and $v$ for a given sample is, to leading exponential order, 
\begin{align}
    \frac{1}{d} \log \mathbb{E}_{x_{\mu}} e^{\hat{u} \langle x+w,x_{\mu}\rangle/\sqrt{2} + \hat{v} \langle x-w,x_{\mu} \rangle/\sqrt{2}} 
    = \zeta(a^2) + o_{d}(1),
\end{align}
where 
\begin{align}
    \zeta(a^2) = \frac{1}{2} \left[ \sqrt{1 + 4 a^2} - 1 - \log\left( \frac{1 + \sqrt{1 + 4 a^2}}{2}\right)\right]
\end{align}
for
\begin{align}
    a^2 &= \frac{1}{d} \left\Vert \hat{u} \frac{x+w}{\sqrt{2}} + \hat{v} \frac{x-w}{\sqrt{2}} \right\Vert^2
    \\
    &= (1+\rho) \hat{u}^2 + (1-\rho) \hat{v}^2 .
\end{align}
Then, we must compute the convex conjugate to obtain the rate function as
\begin{align}
    s(u,v) = \sup_{\hat{u},\hat{v}} \{ u \hat{u} + v \hat{v} - \zeta(a^{2}) \}.
\end{align}
The density of states with a given $u$ and $v$ will then be, to leading exponential order, given by $e^{d [\alpha - s(u,v)]}$ when $u,v$ are such that the exponent is positive; otherwise it is exponentially suppressed. 

Because $u$ and $v$ lie within an ellipse, it is convenient to work in polar coordinates
\begin{align}
    u = \sqrt{1+\rho} r \cos \theta
    \quad \mathrm{and} \quad 
    v = \sqrt{1-\rho} r \sin \theta 
\end{align}
for $r \in [0,1]$ and $\theta \in [0,2\pi)$, in terms of which the rate function reduces to
\begin{align}
    s(r,\theta) = - \frac{1}{2} \log(1-r^2).
\end{align}
Then, the threshold $s(r,\theta) \leq \alpha$ beyond which the density of states becomes suppressed is easily determined, and gives $r = \sqrt{1-e^{-2\alpha}}$. This gives us the asymptotic
\begin{align} \label{eqn:laplace_form}
    \hat{f}_{\mathcal{D}}(x) \sim \frac{\int r\, dr\, d\theta\, e^{d \phi(r,\theta)} g(q(r,\theta))}{\int r\, dr\, d\theta\ e^{d \phi(r,\theta)} }
\end{align}
for a potential 
\begin{align}
    \phi(r,\theta) &= \beta t(r,\theta) + \alpha - s(r,\theta) \\ &= \alpha + \beta r \frac{\sqrt{1+\rho} \cos \theta + \sqrt{1-\rho} \sin \theta}{\sqrt{2}} + \frac{1}{2} \log(1-r^2),
\end{align}
where the integral over $r$ is restricted to $r \leq \sqrt{1-e^{-2\alpha}}$. For $d \to \infty$ the integrals over $r$ and $\theta$ can be evaluated using Laplace's method, which yields the asymptotic
\begin{align}
    \hat{f}_{\mathcal{D}}(x) \sim g(q(r_{\ast},\theta_{\ast})),
\end{align}
where $r_{\ast}$ and $\theta_{\ast}$ maximize $\phi(r,\theta)$ over the allowed region. Maximizing with respect to $\theta$ gives
\begin{align}
    \theta_{\ast} = \arccos \sqrt{{(1+\rho)}/{2}}
\end{align}
for any $r>0$, whereupon the potential for $r$ reduces to
\begin{align} \label{eqn:limiting_potential}
    \phi(r_{\ast},\theta_{\ast}) = \alpha + \beta r_{\ast} + \frac{1}{2} \log(1-r_{\ast}^2)
\end{align}
and $q$ reduces to
\begin{align}
    q(r_{\ast},\theta_{\ast}) = r_{\ast} \frac{\sqrt{1+\rho} \cos \theta_{\ast} - \sqrt{1-\rho} \sin \theta_{\ast}}{\sqrt{2}} = \rho r_{\ast}.
\end{align}
The maximization in $r$ now mirrors the standard REM analysis, hence we have that there exists a phase transition at the `condensation threshold' 
\begin{align}
    \beta_{c} = e^{2 \alpha} \sqrt{1-e^{-2\alpha}} .
\end{align}
In the un-condensed phase $\beta < \beta_{c}$, exponentially many states contribute to the average, and we have 
\begin{align}
    r_{\ast} = \frac{\sqrt{1 + 4 \beta^2}-1}{2\beta} .
\end{align}
In the condensed phase $\beta >\beta_{c}$, the upper limit dominates, and
\begin{align}
    r_{\ast} = \sqrt{1 - e^{-2\alpha}} .
\end{align}
Importantly, $r_{\ast}$ is a non-decreasing continuous function of $\beta$, and is bounded from above by 1. The condensation transition in the REM also has a clear interpretation: if $\beta$ is sufficiently small---\emph{i.e.}, if the kernel bandwidth is sufficiently large---the asymptotic estimate of the function depends only on the bandwidth, not on the total load. However, for very small bandwidths the load becomes important. 

Thus, we at last obtained the claimed asymptotic 
\begin{align} \label{eqn:renormalized_nw}
    \hat{f}_{\mathcal{D}}(x) \sim g(\rho r_{\ast}) .
\end{align}
Therefore, the randomness in the data results in a multiplicative renormalization of $\rho$. In general, generalization improves as $r_{\ast} \uparrow 1$. As a consequence of this renormalization effect, one can obtain an asymptotically unbiased estimator for positive-homogeneous link functions $g$ satisfying $g(\rho r_{\ast}) = r_{\ast}^{k} g(\rho)$ for some $k$ by dividing the NW estimator by $r_{\ast}^{k}$.\footnote{We thank Sabarish Sainathan for this observation.}

\section{Mean-squared generalization error}

The asymptotic \eqref{eqn:renormalized_nw} gives us a prediction for the absolute deviation $|f(x)-\hat{f}_{\mathcal{D}}(x)|$ for a (typical) test point, but one would also want to compute the asymptotics of the mean-squared generalization error. For the single-index models we consider, this reduces to a scalar average over the distribution of $\rho$ induced by the randomness in $x$. Asymptotically, $\rho$ should be approximately Gaussian with variance $1/d$. Given that this distribution is manifestly dimension-dependent, we are confronted with the question of the scale at which we should measure the generalization error, and the question of whether the finite-size corrections to the simple asymptotic for a fixed $x$ must be taken into account in order to accurately predict the mean-squared generalization error. An important effect to consider in this case is the fact that the distribution of fluctuations of the weights in the REM depends strongly on the phase: in the condensed phase, fluctuations become heavy-tailed \cite{derrida1981rem,benarous2005limit}. 

Considering a positive-homogeneous activation of degree $k$, we have the asymptotic $\mathbb{E}_{x}[f(x)^2] = \mathbb{E}_{\rho}[g(\rho)^2] \sim d^{-k} \mathbb{E}_{t \sim \mathcal{N}(0,1)}[g(t)^2]$, which specifies a scale at which to measure the mean-squared error. As this scale is dimension-dependent, contributions from corrections to the pointwise asymptotic may be non-negligible at this scale. Quite generally, we may write
\begin{align}
    \frac{\mathbb{E}_{\mathcal{D}} \mathbb{E}_{x}[ (f(x) - \hat{f}_{\mathcal{D}}(x))^2 ] }{\mathbb{E}_{x}[f(x)^2]} 
    \sim (1-r_{\ast}^{k})^{2} + \delta 
\end{align}
where $\delta = \delta(\alpha,\beta)$ is an error term accounting for potential contributions from corrections, which are conceptually separable from the asymptotic bias term $(1-r_{\ast}^{k})^{2}$. Accurately determining $\delta$ requires a more careful analysis than the rather na\"ive approach followed here, and thus lies outside the scope of this note. We thus leave this result as a conjecture.

\section{Training error}

Considering a typical training point $x_{\nu}$, we have, letting $y_{\mu} = g(\langle w, x_{\mu}\rangle/d)$ for brevity, 
\begin{align}
    \hat{f}_{\mathcal{D}}(x_{\nu}) 
    = \frac{e^{\beta d} y_{\nu} + \sum_{\mu \neq \nu}^{n} e^{\beta \langle x_{\nu}, x_{\mu}\rangle} y_{\mu}}{e^{\beta d} + \sum_{\mu\neq \nu}^{n}  e^{\beta \langle x_{\nu}, x_{\mu}\rangle}} ,
\end{align}
where we have isolated the contribution of the $\nu$-th sample. Now, for large $d$, we can see that there should be two phases: one in which the dominant contribution to $\hat{f}_{\mathcal{D}}(x_{\nu})$ comes from the $\nu$-th sample itself, and one in which the contributions of the remaining $n-1$ samples dominate. The problem of determining when the transition between these two phases occurs is analogous to the problem of determining the capacity of a dense associative memory, as studied by \citet{lucibello2024exponential}. In analogy to the associative memory setting, we refer to the phase in which the $\nu$-th sample dominates the prediction as the retrieval phase. 

\begin{figure}
    \centering
    \includegraphics[height=2in]{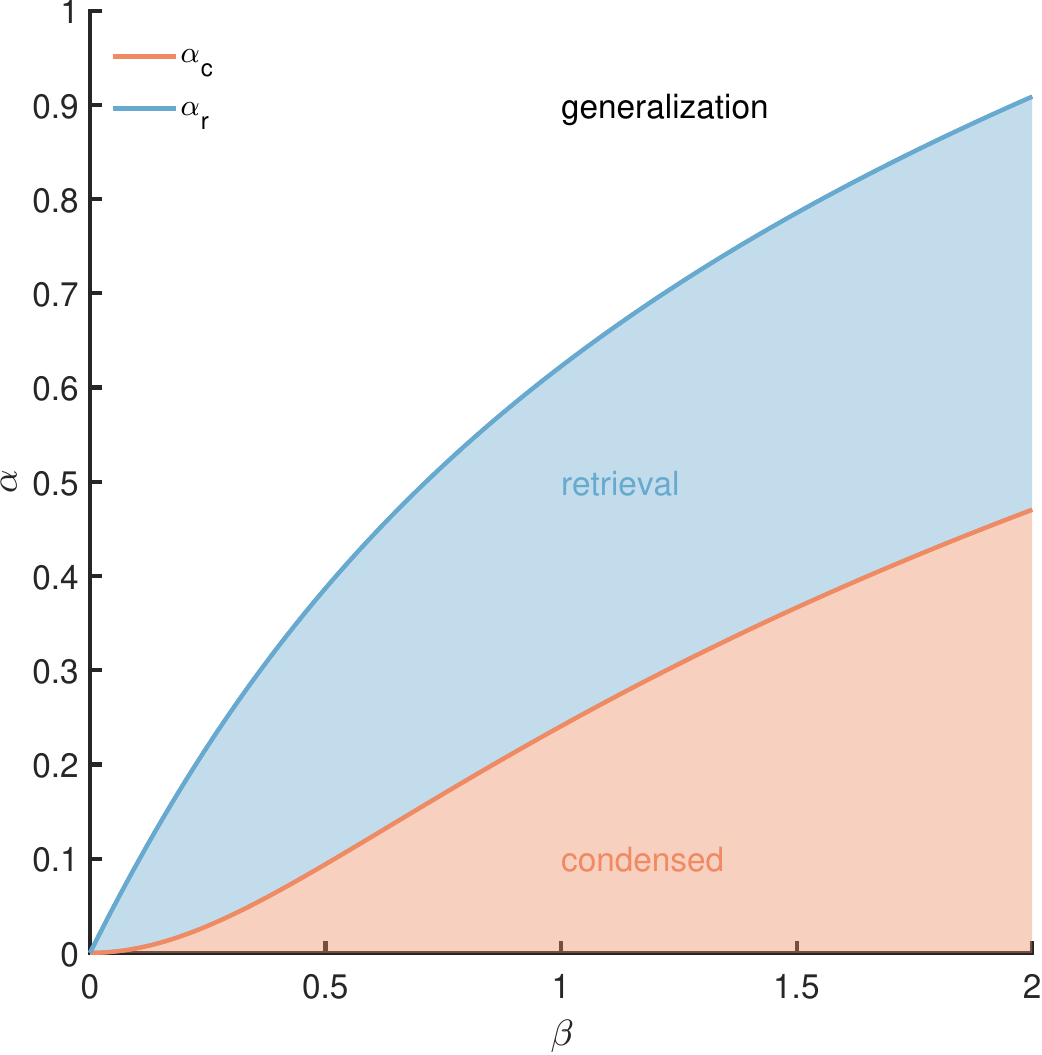}    
    \caption{Phase diagram of the NW estimator in inverse bandwidth $\beta$---load $\alpha$ space. The phase diagram coincides with that of the exponential DAM studied by \citet{lucibello2024exponential}, and the interpretation of the phases is similar. }
    \label{fig:phase_diagram}
\end{figure}

In the retrieval phase, we simply have
\begin{align}
    \hat{f}_{\mathcal{D}}(x_{\nu}) \sim g(\langle w, x_{\nu}\rangle/d) ,
\end{align}
and the training error vanishes. For retrieval to occur, we should have
\begin{align} \textstyle
    \frac{1}{\beta d} \log \sum_{\mu \neq \nu}^{n} e^{\beta \langle x_{\nu}, x_{\mu}\rangle} < 1 ,
\end{align}
which is the condition defining the memorization capacity of a DAM found by \citet{lucibello2024exponential}. The asymptotic behavior of the left-hand-side of this inequality can be obtained using a large deviations computation identical to that above with $q$ integrated out. The resulting condition is that $\phi(r_{\ast},\theta_{\ast}) < \beta$, where $\phi$ is the limiting potential given in \eqref{eqn:limiting_potential}. We then define the retrieval threshold
\begin{align}
    \alpha_{r} = \sup_{\alpha\geq 0} \{ \alpha : \phi(r_{\ast},\theta_{\ast}) < \beta \} .
\end{align}
From the work of \citet{lucibello2024exponential}, we have that in this phase the probability that all spherical patterns may be simultaneously retrieved also tends to one, and that for any fixed $\beta$ the threshold value in $\alpha$ where the retrieval phase terminates appears to be in the non-condensed regime. This leads to a phase diagram which coincides with that for the exponential DAM; we reproduce this in Figure \ref{fig:phase_diagram}. 

Outside of the retrieval phase, the contributions of the $n-1$ other training points to $\hat{f}_{\mathcal{D}}(x_{\nu})$ dominate, and we have
\begin{equation}
    \hat{f}_{\mathcal{D}}(x_{\nu}) 
    \sim  \frac{ \sum_{\mu \neq \nu}^{n} e^{\beta \langle x_{\nu}, x_{\mu}\rangle} y_{\mu}}{\sum_{\mu\neq \nu}^{n}  e^{\beta \langle x_{\nu}, x_{\mu}\rangle}} 
    \sim g\left( r_{\ast} \frac{\langle w, x_{\nu}\rangle}{d} \right)
\end{equation}
conditioned on $x_{\nu}$, by analogy with our study of the prediction on a test point. Then, the computation of the training error matches the computation of the generalization error, so by combining this with the result in the retrieval phase we obtain a complete description of the training performance. In particular, if we again consider positive-homogeneous link functions of degree $k$, we expect that we should have
\begin{align}
    \frac{\frac{1}{n} \sum_{\nu=1}^{n} [f(x_{\nu}) - \hat{f}_{\mathcal{D}}(x_{\nu})]^{2}}{\frac{1}{n} \sum_{\nu=1}^{n} f(x_{\nu})^{2}}
    \sim (1-r_{\ast}^{k})^{2} + \delta
\end{align}
if $\alpha > \alpha_{r}$, subject to the same caveats noted above. The curious feature of this result is then that it predicts a discontinuity in the training error at the retrieval threshold provided that $\delta$ does not smooth the transition too much: either the model can perfectly interpolate all training points, or it incurs a nontrivial relative error matching that on a novel test point. In the phase diagram in Figure \ref{fig:phase_diagram}, we refer to this as the ``generalization'' phase as the model does not distinguish between training and test data. 

\section{Numerical experiments}

\begin{figure*}
    \centering
    \subfloat[\hspace*{1.75in}]{\includegraphics[height=2in]{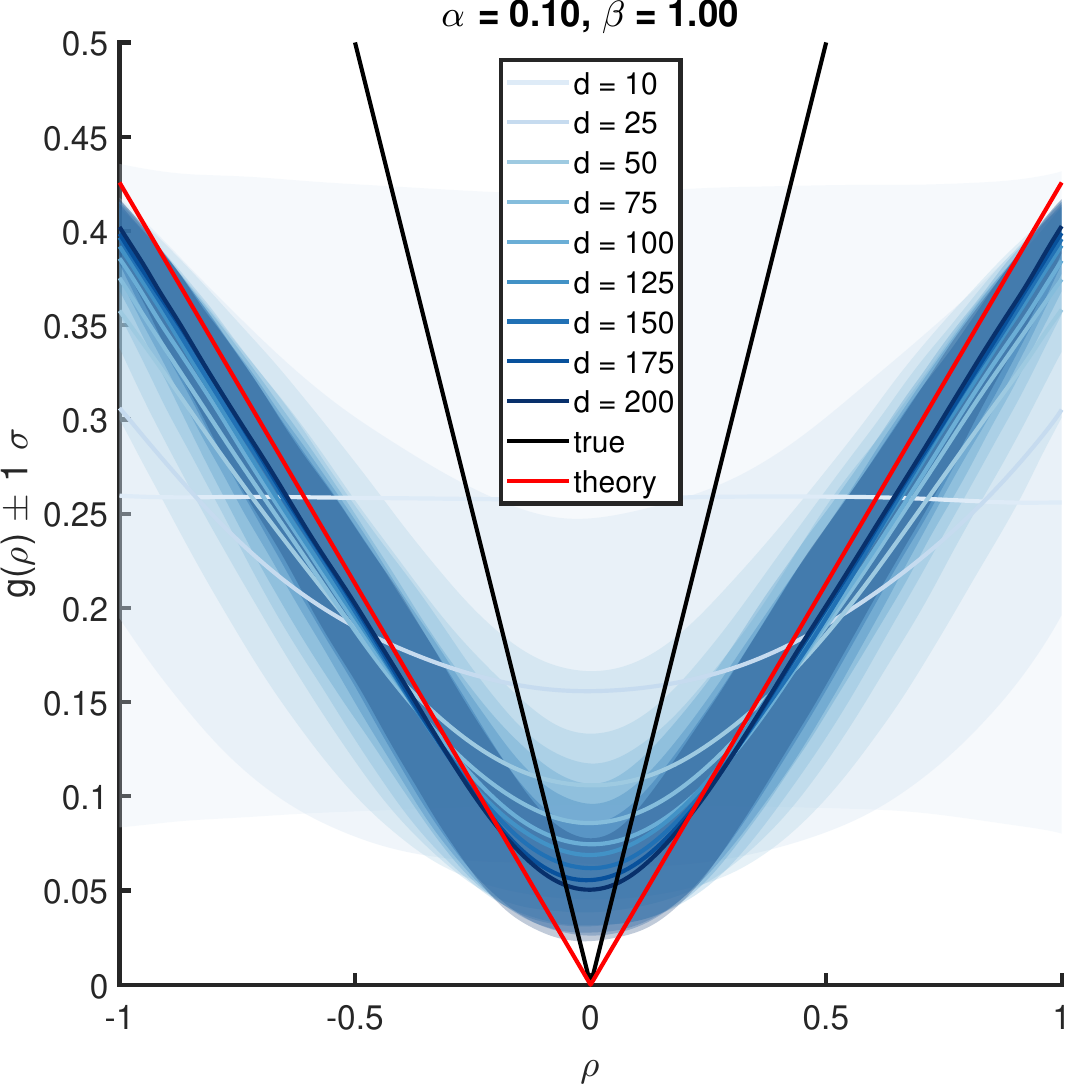}}\hfill%
    \subfloat[\hspace*{1.75in}]{\includegraphics[height=2in]{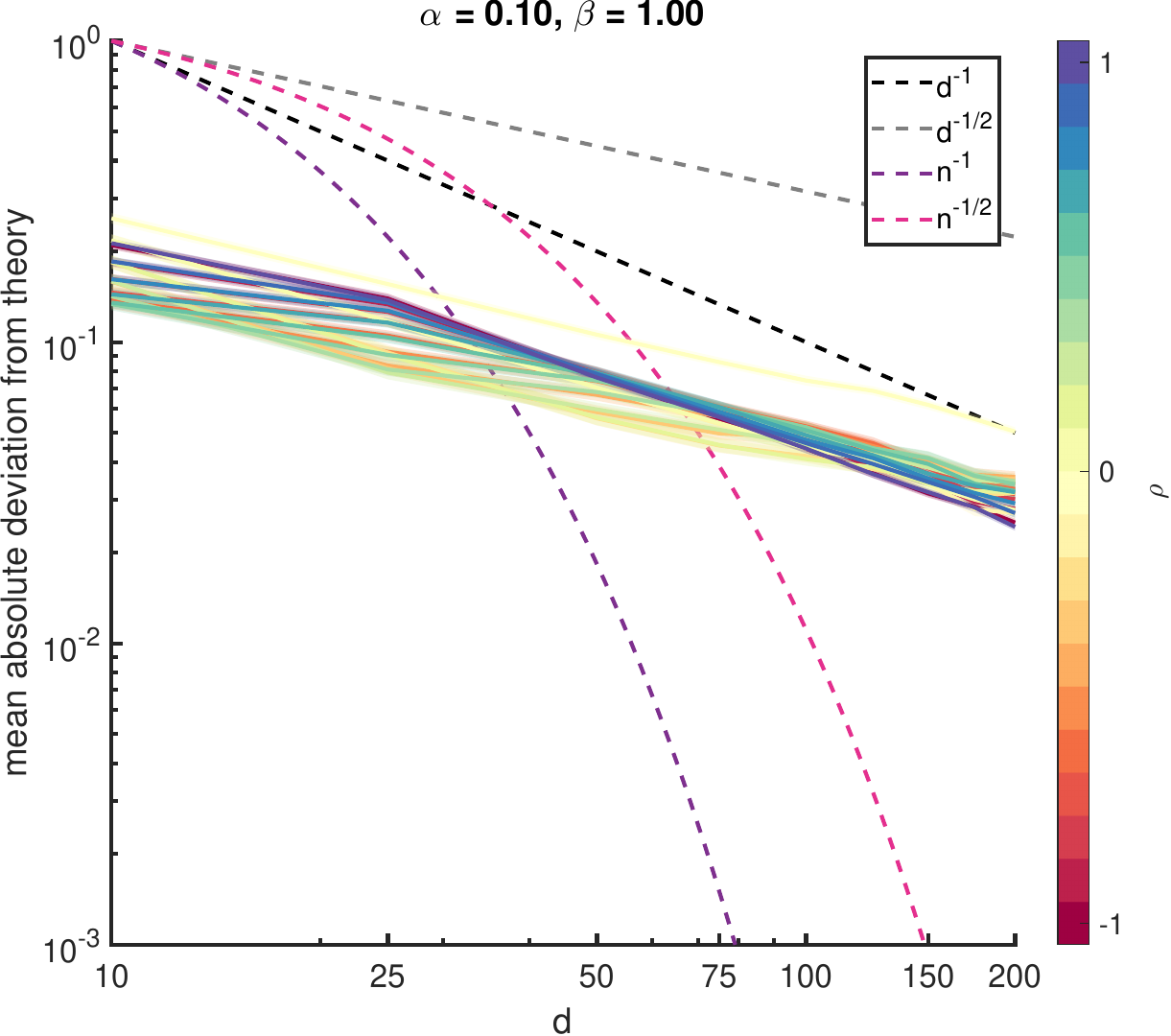}}\hfill%
    \subfloat[\hspace*{1.75in}]{\includegraphics[height=2in]{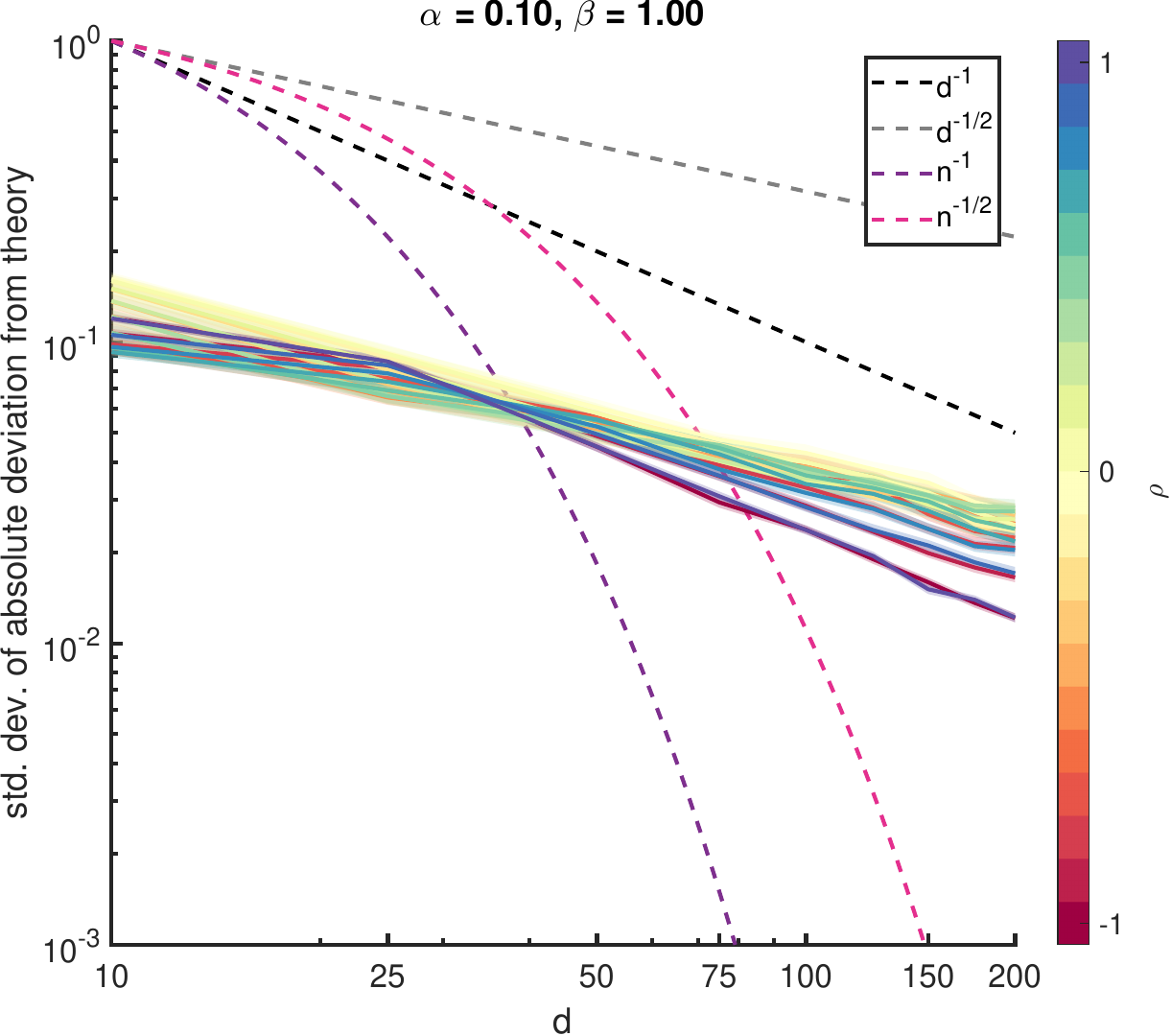}}

    \subfloat[\hspace*{1.75in}]{\includegraphics[height=2in]{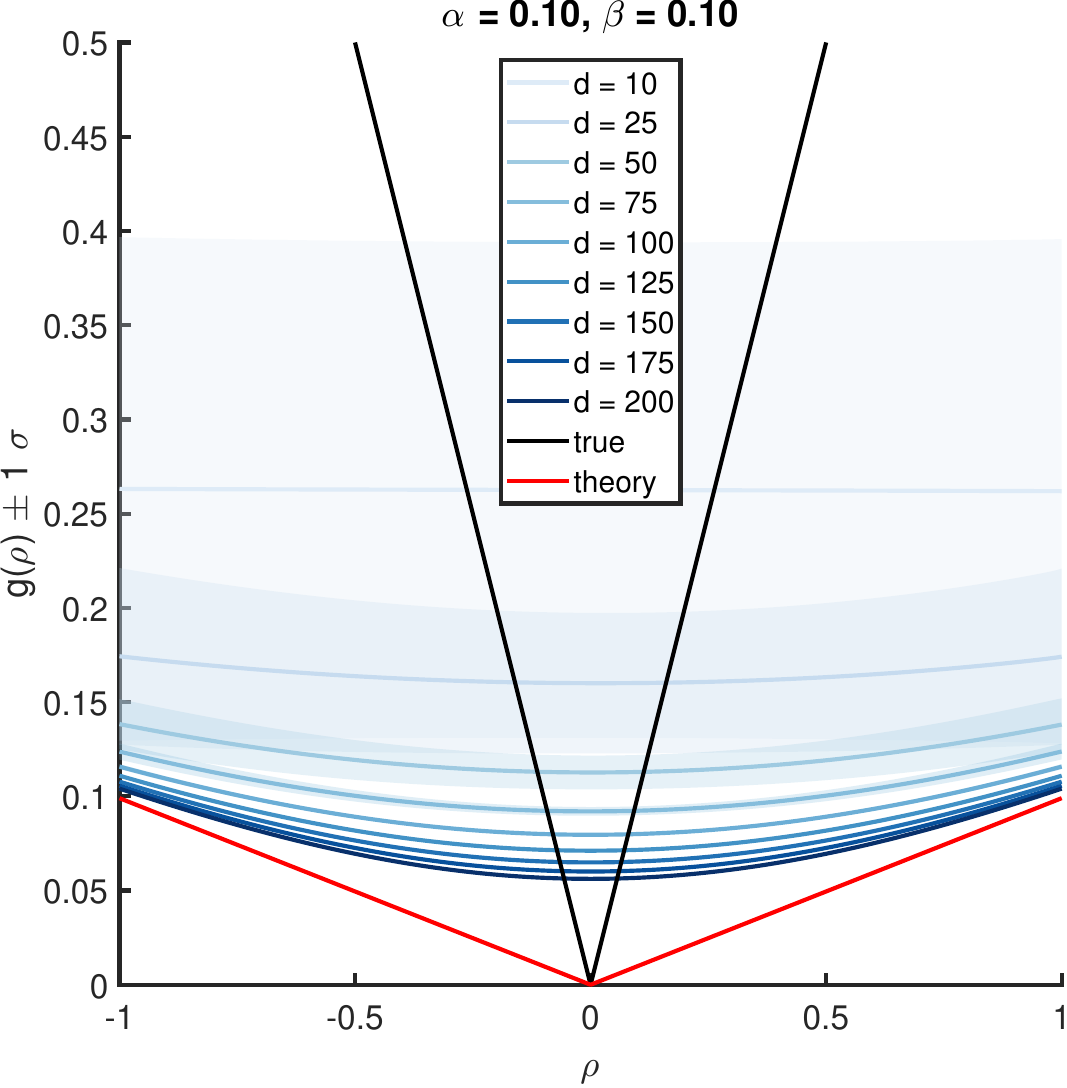}}\hfill%
    \subfloat[\hspace*{1.75in}]{\includegraphics[height=2in]{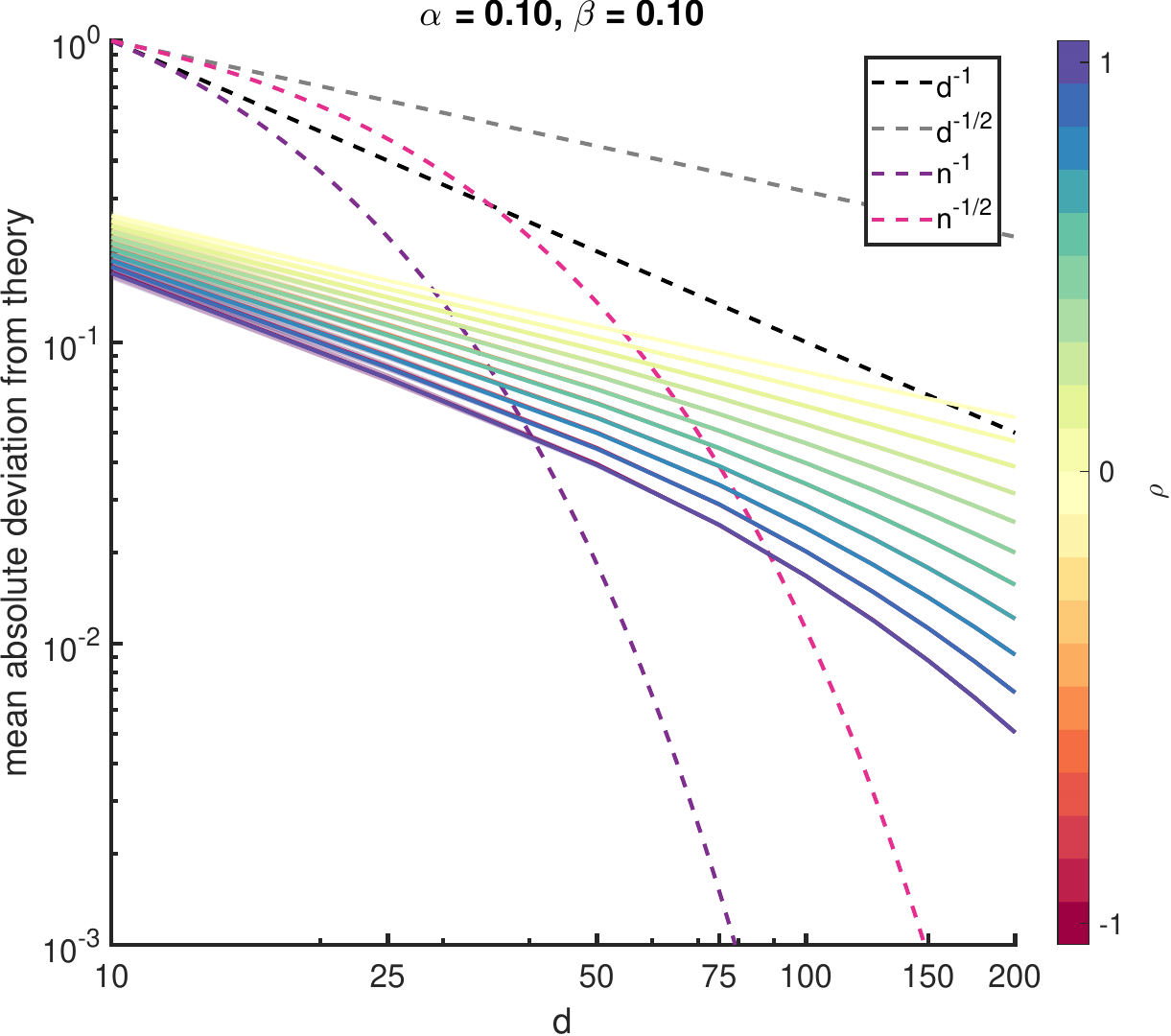}}\hfill%
    \subfloat[\hspace*{1.75in}]{\includegraphics[height=2in]{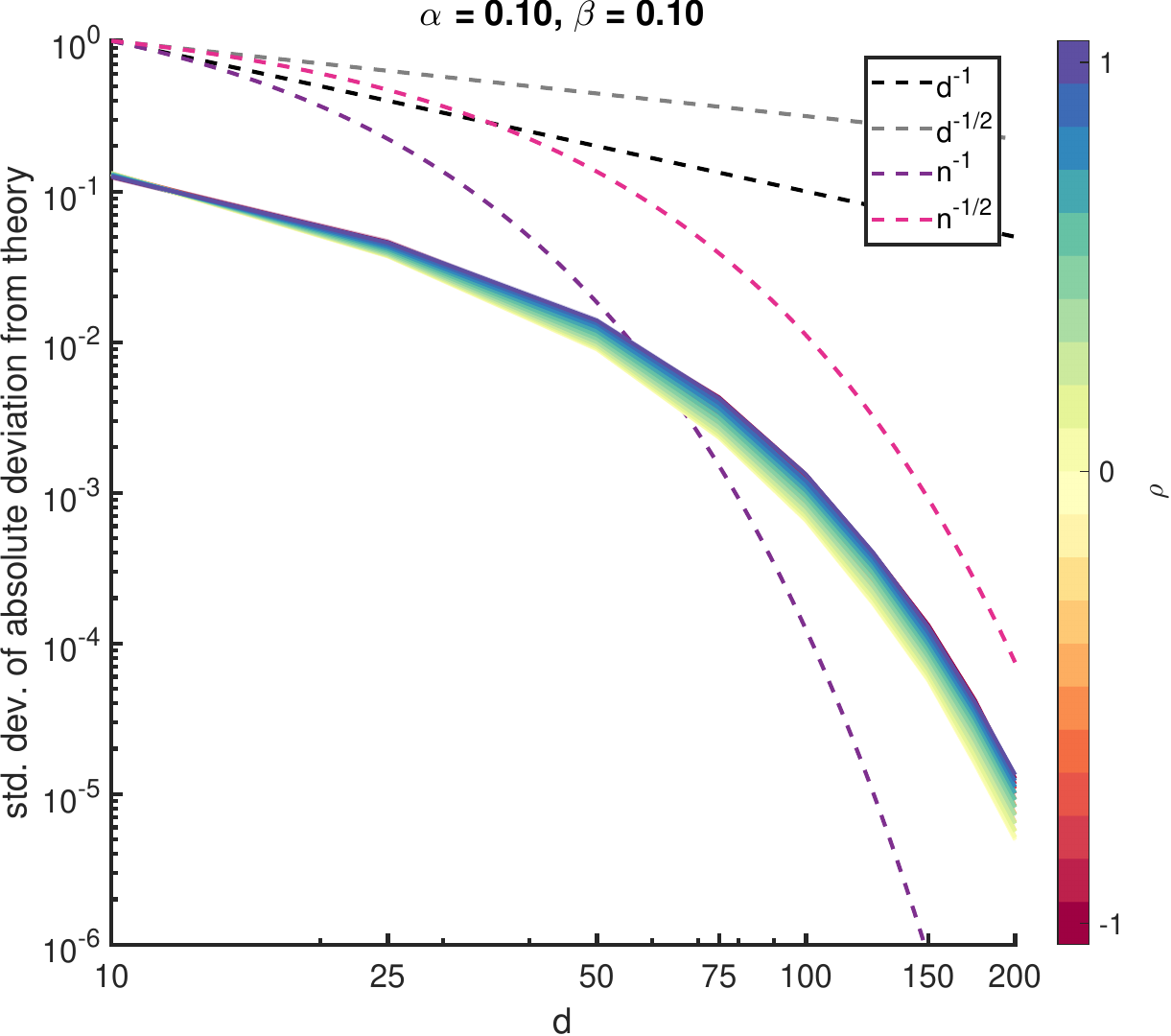}}
    \caption{Simulations of the NW estimator with link function $g(x)=|x|$. (a). Comparison of the asymptotic \eqref{eqn:renormalized_nw} to numerical evaluations of the estimator \eqref{eqn:nw_estimator} in the condensed phase ($\alpha = 0.1$, $\beta = 1$; the condensation threshold is $\beta_{c} = e^{0.2}\sqrt{1-e^{-0.2}} \simeq 0.52$). The red line shows \eqref{eqn:renormalized_nw}, while the black line shows the true link function. Estimates for different dimensions $d$ ranging from 10 to 200 are shown by shades of blue, with shaded patches showing $\pm 1$ standard deviation over 1000 realizations. (b). Mean absolute deviation between numerics and \eqref{eqn:renormalized_nw} as a function of dimension $d$ for different test point overlaps $\rho$ indicated by color. Shaded patches show 95\% confidence intervals computed using the bias-corrected and accelerated percentile bootstrap method. Dashed lines show scalings with $d$ and $p$ to guide the eye. (c). As in (b), but showing the standard deviation of the absolute deviation from \eqref{eqn:renormalized_nw}. (d-f). As in (a-c), but for an example in the un-condensed phase ($\alpha = 0.1$, $\beta = 0.1$). }
    \label{fig:abs}
\end{figure*}

We now would like to compare our theoretical predictions to numerical experiments. The key bottleneck is the exponential dependence of the number of samples on dimension. This makes a na\"ive approach to evaluating the NW estimator impractical due to memory constraints. Instead, we use a slow but memory-efficient iterative algorithm. To do so stably, we multiply and divide \eqref{eqn:nw_estimator} by $e^{-\beta m}/n$, where $m = \max_{\mu} \langle x, x_{\mu} \rangle$ is the maximum dot product, such that both the numerator and denominator of \eqref{eqn:nw_estimator} are bounded. Then, we run the iteration
\begin{equation}
\begin{split} 
    m_{\mu} &= \max\{m_{\mu-1}, \langle x, x_{\mu} \rangle\}
    \\
    Z_{\mu} &= e^{\beta \Delta_{\mu}} Z_{\mu-1} + \frac{1}{\mu} \bigg[ e^{\beta (\langle x,x_{\mu}\rangle - m_{\mu})} - e^{\beta \Delta_{\mu}} Z_{\mu-1}\bigg]
    \\
    \hat{f}_{\mu}(x) &= \frac{e^{\beta \Delta_{\mu}}  Z_{\mu-1} }{Z_{\mu}} \hat{f}_{\mu-1}(x) + \frac{1}{\mu} \left[ \frac{e^{\beta (\langle x, x_{\mu} \rangle-m_{\mu})}}{Z_{\mu}} y_{\mu} - \frac{e^{\beta \Delta_{\mu}} Z_{\mu-1} }{Z_{\mu}} \hat{f}_{\mu-1}(x) \right] 
\end{split}
\end{equation}
with $\Delta_{\mu} = m_{\mu-1} - m_{\mu}$, starting from 
\begin{equation}
\begin{split}
    \hat{f}_{1} &= y_{1}
    \\
    m_{1} &= \langle x, x_{1} \rangle 
    \\
    Z_{1} &= 1 .
\end{split}
\end{equation}
It is easy to check that the endpoint of this iteration gives $\hat{f}_{n}(x) = \hat{f}_{\mathcal{D}}(x)$. As we assume the training examples are independent and identically distributed, we can draw a new datum $x_{\mu}$ at each step, and thus we can avoid instantiating an array containing all $n$ examples. This memory savings comes at the cost of a time complexity linear in $n$---and thus exponential in $d$.

\begin{figure*}
    \centering
    \subfloat[\hspace*{1.75in}]{\includegraphics[height=2in]{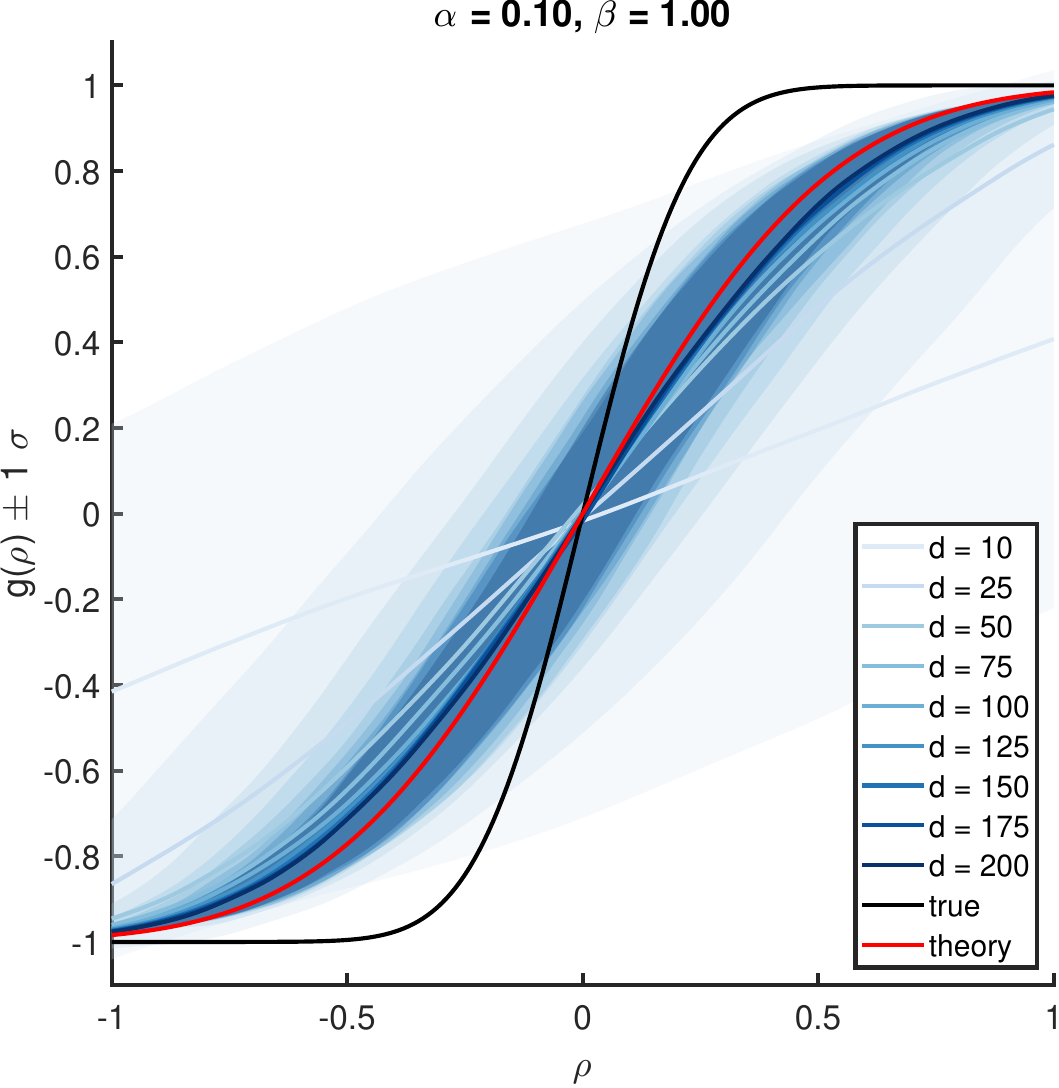}}\hfill%
    \subfloat[\hspace*{1.75in}]{\includegraphics[height=2in]{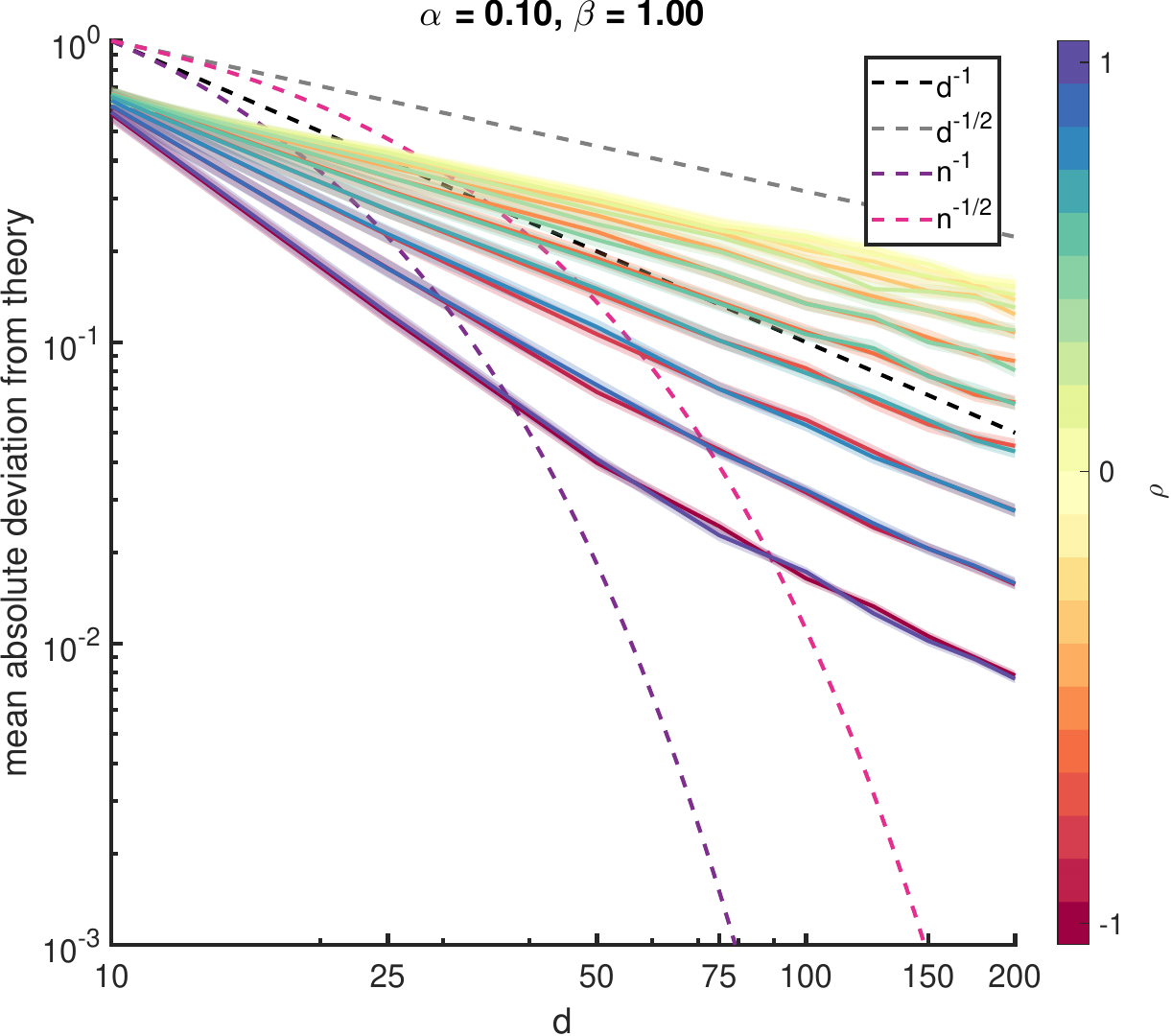}}\hfill%
    \subfloat[\hspace*{1.75in}]{\includegraphics[height=2in]{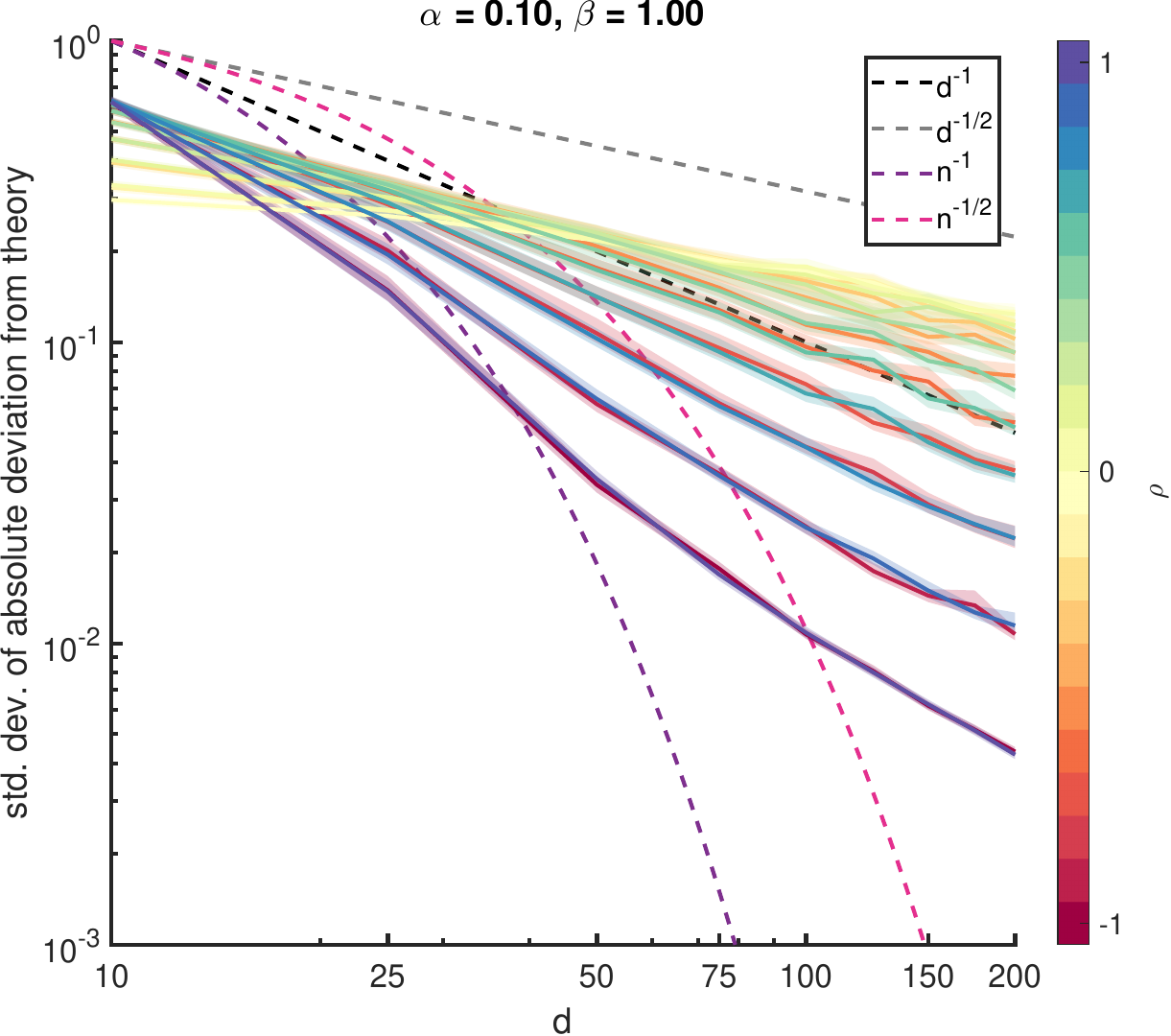}}

    \subfloat[\hspace*{1.75in}]{\includegraphics[height=2in]{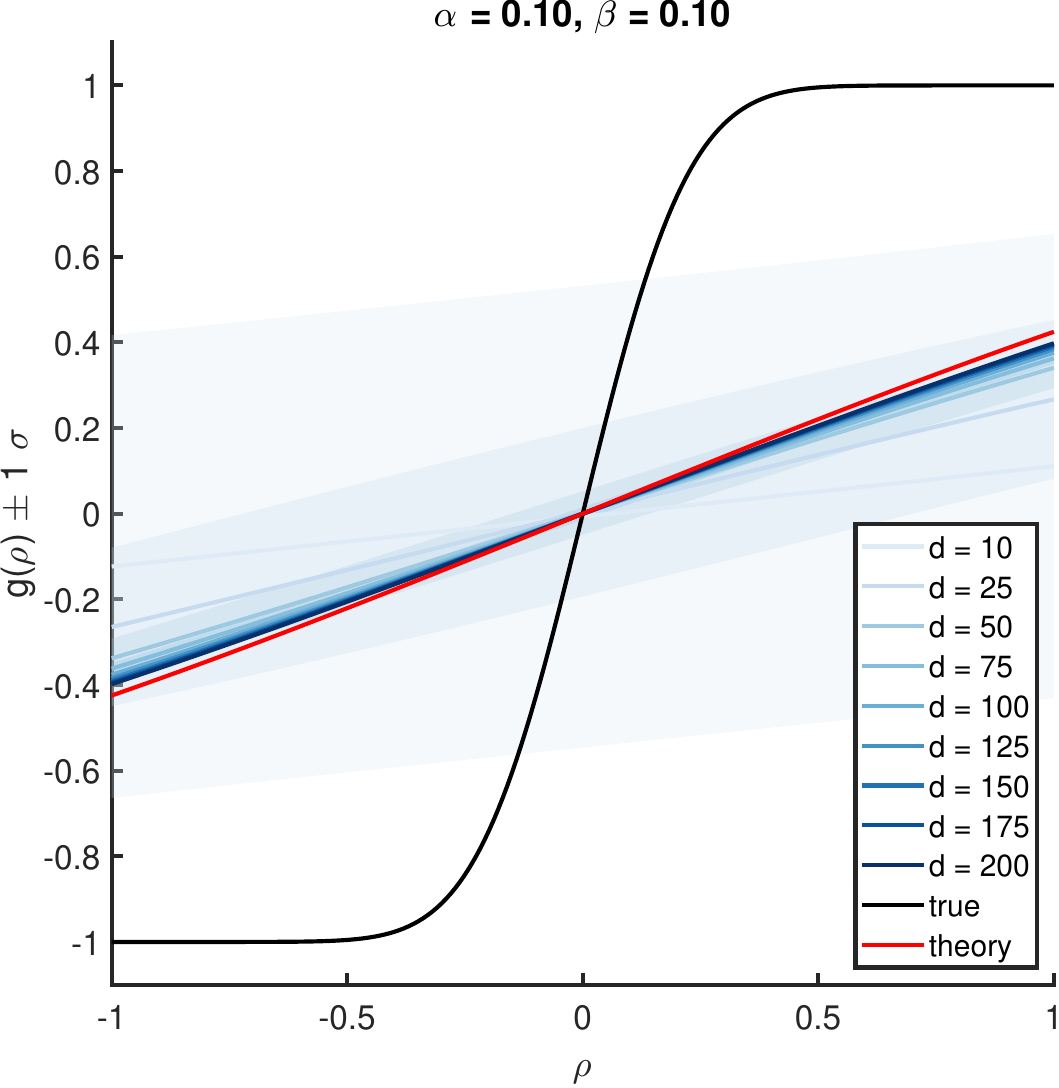}}\hfill%
    \subfloat[\hspace*{1.75in}]{\includegraphics[height=2in]{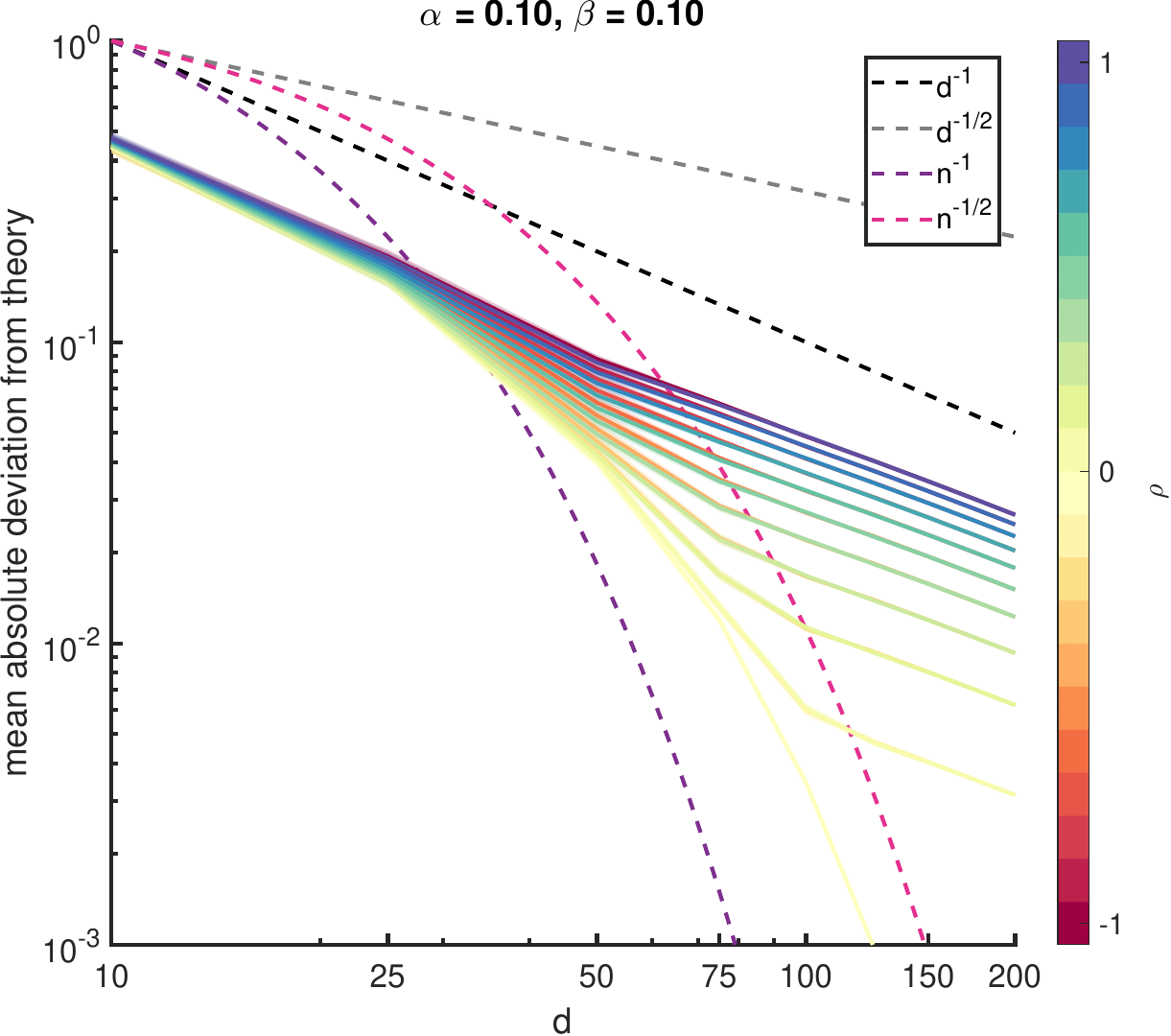}}\hfill%
    \subfloat[\hspace*{1.75in}]{\includegraphics[height=2in]{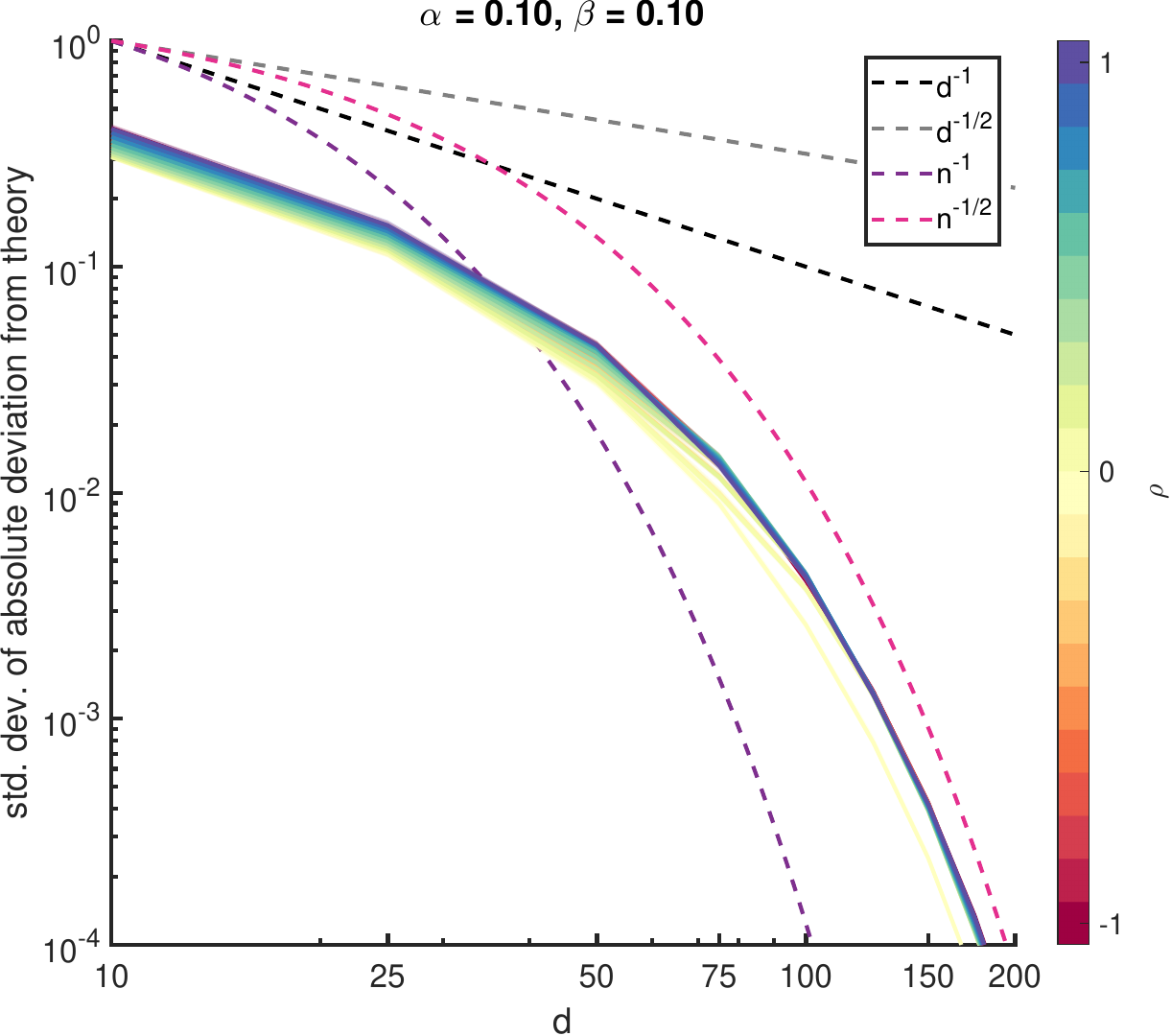}}
    \caption{As in Figure \ref{fig:abs}, but for link function $g(x)=\erf(4x)$. }
    \label{fig:erf}
\end{figure*}

We can now use this algorithm to numerically evaluate the NW estimator for example link functions $g$, and compare the results against the asymptotic predictions obtained in the prelude. Unfortunately, the exponential cost in time is still prohibitive, and thus far we have obtained satisfactory numerical results only for the pointwise prediction of the estimator, not for the training or generalization error. In Figures \ref{fig:abs} and \ref{fig:erf}, we show examples of the predictions for the absolute value function $g(x) = |x|$ and a scaled error function $g(x) = \erf(4x)$, respectively. By examining the size of deviations from the theory, in the condensed phase we appear to have decay of the mean deviation and of the standard deviation at a rate at least $d^{-1/2}$ for $d$ up to 200. In the un-condensed phase, the mean absolute deviation decays at a similar $d^{-1/2}$ rate, while the standard deviation of the absolute deviation increases far more rapidly, seemingly like $n^{-1/2}$. Intuitively, this is to be expected given the number of points that contribute to the estimator in each phase. 

We note that already at this scale we must deal with $e^{20} \simeq 485 \times 10^6$ datapoints, for which each simulation (\textit{i.e.}, a sweep across dimensions for a single link function at a single load and bandwidth) requires around 90 hours of compute time on one 32-core node of Harvard's FASRC Cannon compute cluster. If nothing else, this illustrates the fact that a careful understanding of finite-size effects is required to make predictions about generalization at practically-relevant scales. 

% dimension 1 of 9: 0.848438 s
% dimension 2 of 9: 0.326752 s
% dimension 3 of 9: 0.538981 s
% dimension 4 of 9: 2.021565 s
% dimension 5 of 9: 14.288495 s
% dimension 6 of 9: 142.821616 s
% dimension 7 of 9: 1865.933529 s
% dimension 8 of 9: 24285.466164 s
% dimension 9 of 9: 302679.510457 s

\section{Large target functions}\label{sec:large_target}

Up to this point, we have assumed that the target link function $g$ does not scale with dimension, and thus is small in the sense that it does not affect the maximization of the potential in \eqref{eqn:laplace_form}. If the target function is large enough to change the location of the maximum, the relationship between the true function and the NW estimator is more complicated asymptotically than a simple renormalization of the argument. This can be seen by considering the case in which the target function is itself a radial basis function on the sphere, 
\begin{align}
    g(\langle w,x\rangle/d) = e^{\gamma \langle w,x\rangle},
\end{align}
where $\gamma > 0$ is the inverse bandwidth of the target. Then, to leading exponential order, we are interested in the difference of the free energies of two REMs at different inverse temperatures. Concretely, we consider
\begin{align}
    \frac{1}{d} \log \hat{f}_{\mathcal{D}}(x) = \frac{1}{d} \log \sum_{\mu=1}^{n} e^{\langle \beta x + \gamma w, x_{\mu} \rangle} - \frac{1}{d} \log \sum_{\mu=1}^{n} e^{\beta \langle x, x_{\mu} \rangle} .
\end{align}
From this point, we could either adapt the pointwise large deviations analysis of Section \ref{sec:large_deviations}, or leverage the fact that this is now actually identical to the standard REM computation done in \citet{lucibello2024exponential}. By either route, one finds that
\begin{align}
    \frac{1}{d} \log \hat{f}_{\mathcal{D}}(x) \sim \phi_{\ast}(\sqrt{\beta^2 + \gamma^2 + 2 \beta \gamma \rho}) - \phi_{\ast}(\beta),
\end{align}
where by a minor abuse of notation we write $\phi_{\ast}(\beta) = \phi(r_{\ast}(\beta))$ for the value of the radial potential
\begin{align}
    \phi(r_{\ast}) = \alpha + \beta r_{\ast} + \frac{1}{2} \log(1-r_{\ast}^2) 
\end{align}
maximized with respect to $r \in (0,\sqrt{1-e^{-2\alpha}}]$ for a particular value $\beta$ of the bandwidth. 

This result shows that if the target function is allowed to grow with dimension then the pointwise asymptotic behavior of the NW estimator is not always simple multiplicative renormalization. In particular, this follows from the fact that $\rho$ enters the asymptotic only through $\sqrt{\beta^2 + \gamma^2 + 2 \beta \gamma \rho}$, which unless $\gamma$ is perturbatively small is far from linear in $\rho$. Moreover, in this case the choice of test point can affect whether the estimator is in the condensed phase, which contrasts with what we found in the case of $n$-independent link functions. In particular, if both $\sqrt{\beta^2 + \gamma^2 + 2 \beta \gamma \rho}$ and $\beta$ are above the condensation threshold $e^{2\alpha} \sqrt{1-e^{-2\alpha}}$, then this asymptotic predicts that $\frac{1}{d} \log \hat{f}_{\mathcal{D}}(x)$ will vanish to leading order.

\section{Conclusions}

We have reported the results of a very preliminary investigation of the NW estimator through a random energy model lens. The primary result of our note is the asymptotic \eqref{eqn:renormalized_nw} for the prediction on a fixed test point, which shows that the randomness from the training samples multiplicatively renormalizes the true overlap between the latent vector $w$ and the test point $x$. Our numerical simulations support the accuracy of this prediction, though computational constraints mean that we can access only relatively small dimensions, where finite-size effects are prominent.

There are many problems which we leave open for future inquiry. Though we have contented ourselves with simple asymptotics and comparisons to limited numerics, this analysis could in principle be made entirely quantitative in the sense of explicit error bounds as a function of $d$. Some large deviations results for the NW estimator in fixed dimension $d$ as $n \to \infty$ are known \cite{mokkadem2008largedeviations,tsybakov2008nonparam}, but to our knowledge rigorous results when $d$ and $n$ tend to infinity together are lacking. Given the close relationship of the settings, rigorous bounds on the finite-size effects for the setting considered here would immediately imply similar bounds for memorization of spherical patterns in DAMs \cite{lucibello2024exponential}. 

We have specialized to the setting of spherical data, radial basis function kernels, and a single-index target; in principle all of these assumptions could be relaxed. Extending our results to anisotropic data would be of particular interest, as introducing a preferred direction at some specified angle with the latent vector $w$ would likely alter the condensation threshold due to the local enhancement of data density. A natural starting point is thus to consider data drawn from a von Mises-Fisher distribution, which would require one to track three overlaps between the latent vector $w$, the test point $x$, and the mean direction $\eta$. Extensions to multi-index targets would be of a similar flavor: so long as the number of relevant directions remains finite, one would have to characterize the joint large deviations of the overlap between the input and each direction. This would make the computation more complicated, but at least at the level of a pointwise asymptotic would not introduce new technical challenges. Finally, our approach could be adapted to other kernels provided that one accounts for how their variances scale with dimension. This issue is closely related to classical studies of REM-type models with general energy distributions, as analyzed in detail by \citet{benarous2005limit}. 

Finally, we comment on the broader context of our work. In the last few years, substantial effort has been devoted to seeking precise characterizations of how various learning algorithms behave in high dimensions. As mentioned in the introduction to this note, a key achievement of this program has been a sharp understanding of kernel ridge regression \cite{canatar2021spectral,xiao2022precise,atanasov2024scaling,belkin2019interpolation,spigler2020asymptotic}. It is interesting to contrast the multiplicative renormalization of the argument of the link function observed here to the case of KRR, for which the randomness in the data renormalizes the ridge parameter itself \cite{atanasov2024scaling}. This essay represents a small step towards a similarly detailed understanding of qualitatively different regression algorithms. In concurrent work posted to the arXiv shortly after the initial appearance of this note, \citet{biroli2024kde} have performed a detailed analysis of kernel density estimation that similarly leverages an analogy to the REM. The time is therefore ripe for a full asymptotic description of kernel smoothing methods. Thus, in closing, we echo Jamie Simon's call to arms \cite{simon2024lets}: let's solve more learning rules!

\section*{Acknowledgements}

We thank Jamie Simon for posing the question that inspired this note \cite{simon2024onenn}, and thank Alexander Atanasov, Blake Bordelon, Benjamin Ruben, and especially Sabarish Sainathan for useful discussions. JAZV further thanks Dmitry Krotov for useful discussions regarding \cite{lucibello2024exponential}. 

\paragraph*{Funding acknowledgements} JAZV is supported by a Junior Fellowship from the Harvard Society of Fellows. JAZV and CP were supported by NSF Award DMS-2134157 and NSF CAREER Award IIS-2239780. CP is further supported by a Sloan Research Fellowship. This work has been made possible in part by a gift from the Chan Zuckerberg Initiative Foundation to establish the Kempner Institute for the Study of Natural and Artificial Intelligence. The computations in this paper were run on the FASRC Cannon cluster supported by the FAS Division of Science Research Computing Group at Harvard University.

\paragraph*{Author contributions} JAZ-V conceived the project, performed all research, and wrote the paper. CP supervised the project and contributed to review and editing.

\appendix

\section{Table of notation}\label{sec:notation}

Here, we provide a brief table of notation, with translations to the notation of \citet{lucibello2024exponential} for the reader's convenience. We direct the reader in particular to Appendix C of \citet{lucibello2024exponential}, which contains their analysis of the rate function for spherical data, referred to as ``patterns'' in the context of associative memory models. 
\begin{center}
\begin{tabular}{ c | c | c }
Quantity & Our notation & \citet{lucibello2024exponential} \\ \hline
dimension & $d$ & $N$ \\  
number of training examples (``patterns'') & $n=e^{\alpha d}$ & $P = e^{\alpha N}$ \\
training examples (``patterns'') & $\mathbf{x}_{\mu}$ & $\bm{\xi}^{\mu}$ \\
rate function & $\zeta$ & $\zeta_{\rho}$ \\
REM free energy density & $\beta^{-1} \phi$ & $\phi$ \\
inverse temperature/inverse bandwidth & $\beta$ & $\lambda$
\end{tabular}
\end{center}

\bibliography{refs}

\end{document}